\documentclass[11pt,english,british]{article}
\usepackage{lmodern}
\usepackage[T1]{fontenc}
\usepackage[latin9]{inputenc}
\usepackage{geometry}
\usepackage{color}
\usepackage{amsmath}
\usepackage{amsthm}
\usepackage{amssymb}
\usepackage{soul}
\usepackage{stackrel}
\usepackage{graphicx}
\usepackage{setspace}
\usepackage[authoryear]{natbib}
\makeatletter

\numberwithin{equation}{section}
\numberwithin{figure}{section}

\usepackage{lineno}

\makeatother

\usepackage{babel}
\begin{document}
\title{A Geostatistical Framework for Combining Spatially Referenced Disease Prevalence Data from Multiple Diagnostics}
\author{Benjamin Amoah,
Emanuele Giorgi and
Peter J. Diggle \\
CHICAS, Lancaster University Medical School, Lancaster LA1 4YG, UK}

\date{}

%

\maketitle

\begin{abstract}

Multiple diagnostic tests are often used due to limited resources or because they provide complementary information on the epidemiology of a disease under investigation. Existing statistical methods to combine prevalence data from multiple diagnostics ignore the potential over-dispersion induced by the spatial correlations in the data. To address this issue, we develop a geostatistical framework that allows for joint modelling of data from multiple diagnostics by considering two main classes of inferential problems: (1) to predict prevalence for a gold-standard diagnostic using low-cost and potentially biased alternative tests; (2) to carry out joint prediction of prevalence from multiple tests. We apply the proposed framework to two case studies: mapping \textit{Loa loa} prevalence in Central and West Africa, using miscroscopy and a questionnaire-based test called RAPLOA; mapping \textit{Plasmodium falciparum} malaria prevalence in the highlands of Western Kenya using polymerase chain reaction and a rapid diagnostic test. We also develop a Monte Carlo procedure based on the variogram in order to identify parsimonious geostatistical models that are compatible with the data. Our study highlights ($i$) the importance of accounting for diagnostic-specific residual spatial variation and ($ii$) the benefits accrued from joint geostatistical modelling so as to deliver more reliable and precise inferences on disease prevalence.  

\textbf{\bigskip{}
\\Keywords: }disease mapping; geostatistics;  malaria; neglected tropical disesaes; multiple diagnostic tests; prevalence.
\end{abstract}

\section{Introduction}\label{sec:intro}

\textit{Disease mapping} denotes a class of problems in public health where the scientific goal is to predict the spatial variation in disease risk on a scale that can range from sub-national to global \citep{murray2014global, liu2012global}. Understanding the geographical distribution of a disease is particularly important in the decision-making process for the planning, implementation, monitoring and evaluation of control programmes \citep{world2017world, bhatt2015effect}. In this context, model-based geostatistical methods \citep{diggle1998model} have been especially useful in low-resource settings \citep{diggle2016model, gething2012long, zoure2014geographic} where disease registries are non-existent or geographically incomplete, and monitoring of the disease burden is carried out through cross-sectional surveys and passive surveillance systems. 

It is often the case that prevalence data from a geographical region of interest are obtained using different diagnostic tests for the same disease under investigation. The reasons for this are manifold.   For example, when the goal of geostatistical analysis is to map disease risk on a continental or global scale by combining data from multiple surveys, dealing with the use of different diagnostic tests may be unavoidable. In other cases, gold-standard diagnostic tests are often expensive and require advanced laboratory expertise and technology which may not always be available in constrained resource settings. This requires the use of more cost-effective alternatives for disease testing in order to attain a required sample size. Different diagnostics might also provide complementary information of intrinsic scientific interests into the spatial variation of disease risk and the distribution of hotspots.

In the absence of statistical methods that allow for the joint analysis of multiple diagnostics, most studies have reported separate analyses. A shortcoming of this approach is that it ignores, and therefore fails to explain, possible correlations between prevalence of different diagnostics. Statistical inference might benefit from a joint analysis, which can yield more efficient estimation of regression parameters \citep{song2009joint} and more precise predictions of prevalence.

However, different diagnostic tests can exhibit considerable disparities in the estimates of disease prevalence for the same population, or even the same individuals. Obvious sources of such variation include differences
in sensitivity and specificity. Furthermore, different diagnostics may exhibit differences in their association with the same risk factors. In a geostatistical context, there may also be differences between the spatial covariance structures of different diagnostics.

These aspects highlight the potential challenges that joint modelling of multiple diagnostics needs to take into account. In this paper, we address such issues in order to develop a general framework for geostatistical analysis and describe the application of this framework
to \textit{Loa loa} and malaria mapping in Africa.

The structure of the paper is as follows. In Section \ref{sec:problem}, we describe the two motivating applications. In Section \ref{sec:literature_review}, we review existing methods for combining prevalence data from different diagnostics. 
In Section \ref{sec:modelling}, we introduce a geostatistical framework for combining data from two diagnostics and distinguish two main classes of problems that arise in this context. In Sections \ref{sec:application} and \ref{sec:application2}, we apply this framework to the two case studies introduced in Section \ref{sec:problem}. In Section \ref{sec:discussion}, we discuss methodological extensions to more than two diagnostics. 

\section{Motivating applications}\label{sec:problem}

\subsection{Loa loa mapping in Central and West Africa}\label{sec:exam_case1}

Loiasis is a neglected tropical disease that has received an increased attention due to its impact on the control of a more serious infectious disease, onchocerciasis, that is endemic in large swathes of sub-Saharan Africa. Mass administration of
the  drug ivermectin confers protection against onchocerciasis,
 but individuals who are highly co-infected with \textit{Loa loa} - the Loiasis parasite - can develop severe and occasionally fatal adverse reaction to the drug \citep{boussinesq1998three}. 

\citet{boussinesq2001relationships} have shown that high levels in \textit{Loa loa} prevalence within a community are strongly associated with a high parasite density. For this reason, \citet{zoure2011} have suggested that precautionary measures should be put in place before the roll-out of mass drug administration with ivermectin in areas where prevalence 
of infection with \textit{Loa loa} is greater than 20\%.

In order to carry out a rapid assessment of the \textit{Loa loa} burden in endemic areas a questionnaire instrument, named RAPLOA, was developed as a more economically feasible alternative to the standard microscopy-based microfilariae (MF) detection in blood smears \citep{takougang2002rapid}. To validate the RAPLOA methodology against microscopy,  cross-sectional surveys using both diagnostics were carried out in four study sites in Cameroon, Nigeria, Republic of Congo and the Democratic Republic of Congo (see \cite{wanji2012validation} and Additional Figure 1 in Web Appendix B).

In this study, the objective of statistical inference is to develop a calibration relationship between the two diagnostic procedures. This could then be applied to map microscopy-based MF prevalence in areas where the more economical RAPLOA questionnaire is the only feasible option.

\subsection{Malaria mapping in the highlands of Western Kenya}\label{sec:exam_case2}

Malaria continues to be a global public health challenge, especially in sub-Sharan Africa which, in 2016, accounted for about 90\% of all the 
445,000 estimated malaria deaths worldwide \citep{world2017world}. Polymerase chain reaction (PCR) and a rapid diagnostic test (RDT) are two of the most commonly used procedures for detecting \textit{Plasmodium falciparum}, the deadliest species of the malaria parasites.  PCR is highly sensitive and specific, but its use is constrained by high costs and the need for highly trained technicians. RDT is simpler to use, cost-effective and requires minimal training, but is less sensitive than PCR \citep{tangpukdee2009malaria}. Recent studies have reported that PCR and RDT can lead to the identification of different malaria hotspots, i.e. areas where disease risk is estimated to be unexpectedly high  \citep{mogeni2017detecting}.  In this context, mapping of both diagnostics is of epidemiological interest since their effective use is dependent on the level of malaria transmission, with PCR being the preferred testing option in low-transmission settings  \citep{mogeni2017detecting}.

In order to investigate this issue, a malariometric survey was conducted using both RDT and PCR in two  highland districts of Western Kenya (see Additional Figure 2 in Web Appendix B); see \citet{stevenson2015use} for a descriptive analysis of this study. In this scenario, a joint model for the reported malaria counts from the two diagnostics could allow to exploit their cross-correlation and identify malaria hotspots more accurately.

\section{Literature review}\label{sec:literature_review}

We formally express the format of geostatistical data from multiple diagnostics as
\begin{equation}
\label{eq:data1}
\mathcal{D} = \{ (x_{ik}, n_{ik}, y_{ijk}): j=1,\ldots,n_{ik};i=1,\ldots,N;k=1,\ldots,K \}
\end{equation}
where $y_{ijk}$ is a binary outcome taking value 1 if the $j$-th individual at location $x_{ik}$ tests positive for the disease under investigation using the $k$-th diagnostic procedure, and 0 otherwise. 
We use $p_{ijk}$ to denote the probability that an individual has a positive test outcome from the $k$-th diagnostic. When data are only available as aggregated counts, we replace $y_{ijk}$ in \eqref{eq:data1} with $y_{ik} = \sum_{j=1}^{n_{ik}} y_{ijk}$ and $p_{ijk}$ with $p_{ik}$. When all diagnostic tools are used at each location, we replace $x_{ik}$ with $x_{i}$, although this is not a requirement in the development of our methodology.

In the remainder of this section, we review non-spatial methods for joint modelling of the $p_{ijk}$ across multiple diagnostics and a geostatistical modelling approach proposed by \citet{crainiceanu2008bivariate}. 

\subsection{Non-spatial approaches}
Existing non-spatial methods for the analysis of data from multiple diagnostics fall within two main classes of statistical models: generalised linear models (GLMs) and their random-effects counterpart, generalised linear mixed models (GLMMs).

\cite{mappin2015standardizing} analysed data on \textit{P. falciparum} prevalence from RDT and microscopy outcomes from sub-Saharan Africa, using a  standard probit model
\begin{eqnarray}
\label{eq:probit_mappin}
\Phi^{-1} (p_{i1}) = \beta_0 + \beta_1 \Phi^{-1} (p_{i2}), 
\end{eqnarray}
thus assuming a linear relationship  between the $p_{ik}$ on the probit scale. \cite{wanji2012validation} used a similar  approach for \textit{Loa loa} in order to study the relationship between microscopy and RAPLOA prevalence by replacing the probit link in \eqref{eq:probit_mappin} with the logit. This model was also used by \cite{wu2015comparison} to estimate the relationship between RDT, microscopy and PCR, for each pair of diagnostics. A major limitation of these approaches based on standard GLMs is that they do not account for any over-dispersion that might be induced by unmeasured risk factors. 

\cite{coffeng2013onchocerciasis} proposes a bivariate GLMM for joint modelling of data on onchocerciasis nodule prevalence and skin MF prevalence in adult males sampled across 148 villages in 16 African countries. More specifically, the linear predictor of such model can be expressed as
\begin{eqnarray}
\log \left\{\frac{p_{ijk}}{1-p_{ijk}}\right\} = d_{ij}^\top \beta_k + Z_i+ V_{ij} , 
\end{eqnarray}
where the random effects terms $Z_i$ and $V_{ij}$ are zero-mean Gaussian variables accounting for unexplained variation
between-villages and between-individuals within villages, respectively. Using this approach, \cite{coffeng2013onchocerciasis} estimated a strong positive correlation between nodule and MF prevalence but also reported a variation in the strength of this relationship across study sites.

\subsection{The Crainiceanu, Diggle and Rowlingson model}\label{sec:loaloa_review}
\cite{crainiceanu2008bivariate} proposed a bivariate  geostatistical model (henceforth CDRM) to analyse data on microscopy and RAPLOA \textit{Loa loa} prevalence (see Section \ref{sec:exam_case1}). To the best of our knowledge, this is the
only existing approach that attempts to model the spatial correlation between two diagnostics. 

Let $k=1$ correspond to the RAPLOA questionnaire, and $k=2$ to microscopy. To emphasize the spatial context, we now write $p_{ij}=p_{j}(x_i)$; CDRM can then be expressed as
\begin{eqnarray}\label{eqn:diggle}
\begin{cases}
\text{logit}\{p_1(x_i)\} = d^\top(x_i)\beta + S(x_i) 		\\
\text{logit}\{p_2(x_i)\} = \alpha_0 + \alpha_1 \text{logit}\{p_1(x_i)\} + Z_i , 
\end{cases}
\end{eqnarray}
where ${\rm logit}(u)=\log\{u/[1-u]\}$, $d(x_i)$ is a vector of spatially varying covariates, $S(x_i)$ is a zero-mean stationary and isotropic Gaussian process and the $Z_i$ are zero-mean independent and identically distributed Gaussian random variables. \cite{crainiceanu2008bivariate} also provide empirical evidence to justify the assumption of a logit-linear relationship between the two diagnostics. 

A limitation of the CDRM is that it assumes proportionality on the logit scale between the residual spatial fields associated with RAPLOA and microscopy. In our re-analysis in Section \ref{sec:application}, we use a Monte Carlo procedure to test this hypothesis. 

\section{Two classes of bivariate geostatistical models}\label{sec:modelling}

We now develop two modelling strategies that address the specific objectives of the two case studies introduced in Section \ref{sec:problem}. Our focus in this section will be restricted to the case of two diagnostics (hence $K=2$). We discuss the extension to more than two in Section \ref{sec:discussion}.

\subsection{Case I: Predicting prevalence for a gold-standard diagnostic}
\label{sec:model_formulation1}

Let $S_{1}(x)$ and $S_{2}(x)$ be two independent stationary and isotropic Gaussian processes; also, let $f_{1}\{\cdot\}$ and $f_{2}\{\cdot\}$ be two functions with domain on the unit interval $[0,1]$ and image on the real line. We propose to model data from two diagnostics, with $k=2$ denoting the gold-standard, as
\begin{eqnarray}\label{eqn:mod}
\begin{cases}
f_1\{p_{1}(x_i)\} = d^\top(x_i)\beta_1 + S_1(x_i) + Z_{i1}		\\
f_2\{p_{2}(x_i)\} = d^\top(x_i)\beta_2 + S_2(x_i) + Z_{i2}  + \alpha f_1\{p_{1}(x_i)\}.
\end{cases}
\end{eqnarray}

In our applications, we
specify exponential correlation functions for $S_k(x)$, $k=1, 2$, hence
$$
{\rm cov}\{S_{k}(x), S_{k}(x')\} = \sigma_{k}^2 \exp\{\|x-x'\|/\phi_{k}\},
$$
where $\sigma^2_k$ is the variance of $S_{k}(x)$ and $\phi_k$ is a scale parameter regulating how fast the spatial correlation decays to zero for increasing distance. Finally, we use $\tau^2_k$ to denote the variance of the Gaussian noise $Z_{ik}$.

Selection of suitable functions $f_{1}$ and $f_{2}$ can be carried out, for example, by exploring the association between the empirical prevalences of the two diagnostics in order to identify transformations that render their relationship approximately linear. Alternatively, subject matter knowledge could be used to constrain the admissible forms for $f_{1}$ and $f_{2}$; see, for example, \cite{irvine2016understanding} who derive a functional relationship between MF and an immuno-chromatographic test for prevalence of lymphatic filariasis by making explicit assumptions on the distribution of worms and their reproductive rate in the general population.

The proposed model in \eqref{eqn:mod} is more flexible than the CDRM because ($i$) it allows for diagnostic-specific unstructured variation $Z_{ik}$ and, more importantly, ($ii$) relaxes the assumption of proportionality between the residual spatial fields of the two diagnostics through the introduction of $S_{2}(x)$.   

\subsection{Case II: Joint prediction of prevalence from two complementary diagnostics}
\label{sec:model_formulation2}
Let $S_{1}(x)$ and $S_{2}(x)$ be two independent Gaussian processes, and $Z_{ik}$ Gaussian noise, each having the same properties as defined in the previous section. We now introduce a third stationary and  isotropic Gaussian process $T(x)$ having unit variance and exponential correlation function with scale parameter $\phi_T$.

Our proposed approach for joint prediction of prevalence from two diagnostics, when both are of interest, is expressed by the following equation for the linear predictor
\begin{eqnarray}\label{eqn:mod_mal}
f_k\{p_{jk}(x_i)\} = d_{ij}^\top\beta_k + \nu_k\big[ S_k(x_i) + T(x_i) \big] + Z_{ik}.
\end{eqnarray}
The spatial processes $S_k(x)$ and $T(x)$ accounts for unmeasured risk factors that are specific to each and common to both diagnostics, respectively. The resulting variogram for the linear predictor is
\begin{eqnarray}
\gamma_k(u) &=& \text{\Large{E}} \Big[ \Big\{\Big( \nu_k   	\big( S_k(x) + T(x) \big) + Z_k(x) \Big) - \Big( \nu_k   	\big( S_k(x') + T(x') \big) + Z_k(x') \Big)\Big\}^2 \Big] \nonumber\\ 
					&=& \tau_k^2 + \nu_k^2 \big[1-  \exp({-u}/{\phi_{T}}\big) + \sigma_k^2\big\{1- \exp \big({-u}/{\phi_{S_k}}\big) \big\} \big], 
\end{eqnarray}
and the cross-variogram between the linear predictors of the two diagnostics is 
\begin{eqnarray}
\gamma_{1,2}(u) &=& \text{\Large{E}} \Big[ \Big\{\Big( \nu_1   	\big( S_1(x) + T(x) \big) + Z_k(x)\Big) - \Big(\nu_2   	\big( S_2(x') + T(x') \big) + Z_k(x') \Big)\Big\}^2 \Big] \nonumber\\ 
					&=& 0.5\{\tau_1^2 + \tau_2^2 + \nu_1^2( 1 + \sigma_1^2) + \nu_2^2(1+\sigma_2^2) \} - \sigma_1 \sigma_2 \exp({-u}/{\phi_{T}}).
\end{eqnarray}

Given the relatively large number of parameters,
fitting the model may require a pragmatic approach. In order to identify a parsimonious model for the data, we recommend an incremental modelling strategy, whereby a simpler model is used in a first analysis (e.g. by setting $S_{k}(x)=0$ for all $x$) and more complexity is then added in response to an unsatisfactory validation check, as described below.

\subsection{Comparison between the two models}
\label{subsec:comparison}
\begin{figure}
 \centerline{\includegraphics[width=6.3in]{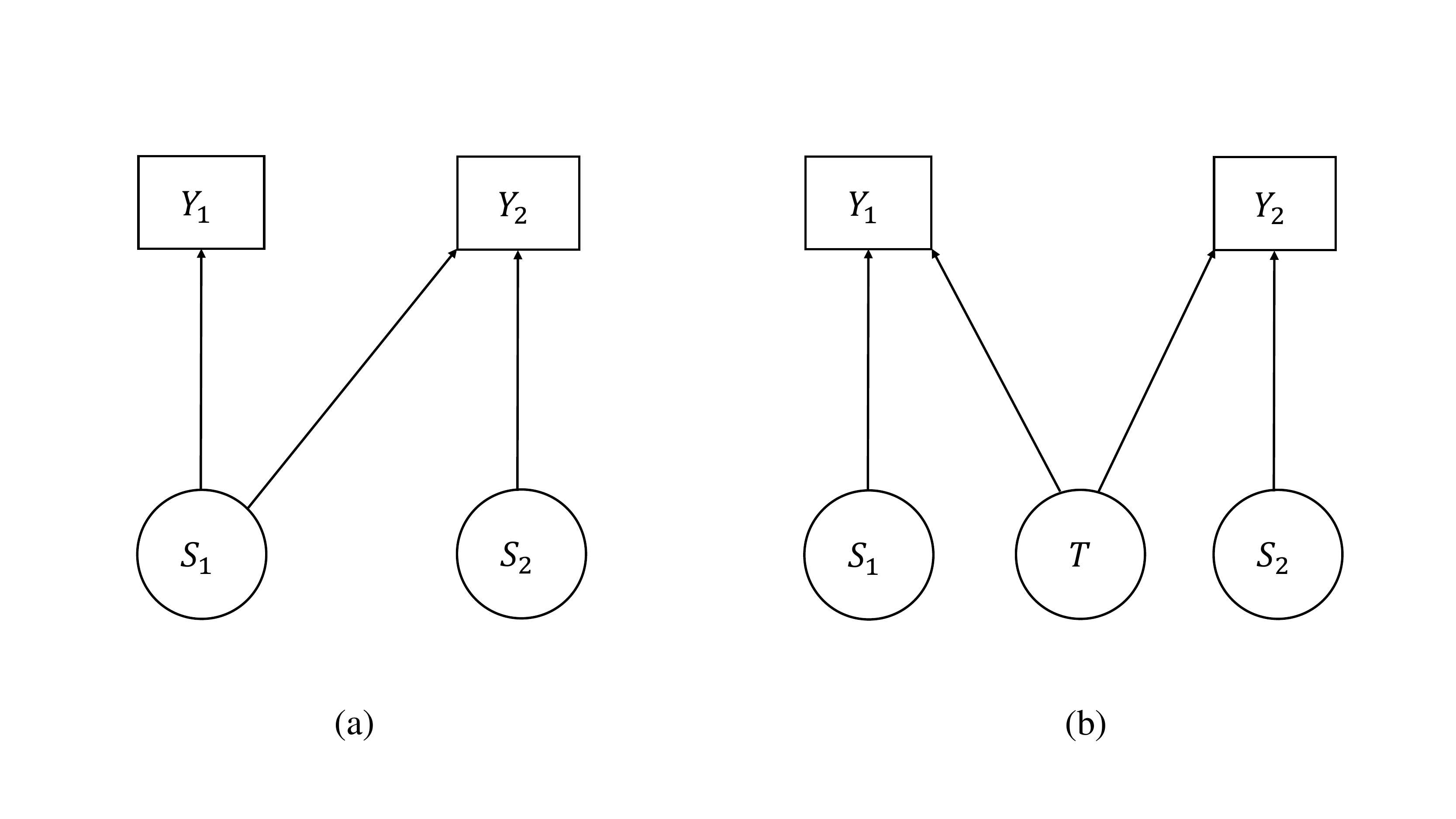}}
\caption{Directed acyclic graphs for the bivariate geostatistical models in \eqref{eqn:mod} (left panel) and \eqref{eqn:mod_mal} (right panel). Circles and squares identify latent variables and the outcome random variables, respectively.}
\label{fig:models}
\end{figure}

Figure \ref{fig:models} gives two directed acyclic graph representations of the models in \eqref{eqn:mod} (left panel) and \eqref{eqn:mod_mal} (right panel), showing their distinctive asymmetric and symmetric structures. In the first model, stochastic independence between the two diagnostics is simply recovered by setting the parameter $\alpha=0$.
If this is a scientifically relevant hypothesis, we can test
it through the likelihood ratio. In the second model, independence can only be achieved if $T(x)=0$ for all $x$. We do not consider 
this to be a credible assumption for the malaria application of Section \ref{sec:exam_case1}.

\subsection{Inference and model validation}\label{sec:inferences}

We carry out parameter estimation for both the asymmetric and symmetric models using Monte Carlo Maximum Likelihood (MCML)  \citep{geyer1991markov, christensen2004monte}. To carry out spatial predictions at a set of unobserved locations, we plug the MCML estimates into a Markov Chain Monte Carlo algorithm for simulation from the distribution of the random effects conditional on the data. We summarise our predictive inferences on prevalence using the mean, standard deviation, and exceedance probabilities, i.e. the probability that the predictive distribution of prevalence exceeds a predefined threshold. Details of the derivation and approximation of the log-likelihood function are given in Web Appendix A. 

For model validation we propose the following procedure. We first re-write both models in general form
\begin{eqnarray}\label{eqn_gstandard} 
f_k\{p_{jk}(x_i)\} = \mu_{ijk} + W_k(x_i),  
\end{eqnarray}
where $\mu_{ijk}$ is the mean component expressed as a regression on the available covariates. In (\ref{eqn_gstandard}), if we set $W_1(x_i) = S_1(x) + Z_{i1}$ and $W_2(x_i) = S_2(x_i) + Z_{i2}  + \alpha \{f_{1}(x_i)\}$, then \eqref{eqn_gstandard} reduces to the asymmetric model \eqref{eqn:mod}; if, instead, $W_k(x_i) =\nu_k\big( S_k(x_i) + T(x_i) \big) + Z_{ik}$, we recover the symmetric model \eqref{eqn:mod_mal}.  

We define the empirical variogram of $W_k(x)$ to be
\begin{eqnarray}
\label{eq:empir_variog}
\hat{\gamma}_k(u) = \frac{1}{2|N(u)|} \sum \limits_{(i,j) \in N(u)} \big\{ \hat{W}_k(x_i)-\hat{W}_k(x_j) \big\}^2,
\end{eqnarray} 
where $N(u) = \{ (i, j) : ||x_i - x_j|| = u, i \neq j \}$ and $\hat{W}_{k}(x_{i})$ is the mean of distribution of $W_{k}(x_{i})$ conditioned to the data. To test whether the adopted spatial structure for $W_{k}(x)$ is compatible with the data, we then proceed through the following steps.

\begin{enumerate}
\item[\textit{Step 0.}] Obtain $\hat{W}_k(x_i)$ from two separate standard geostatistical models (i.e. $W_{k}(x_{i})=S_{k}(x_{i})+Z_{ik}$, where $S_{k}(x)$, $k=1,2$ are independent processes) and compute the empirical variogram $\hat{\gamma}_{k}$, for $k=1,2$.
\item[\textit{Step 1.}] Simulate prevalence data as in \eqref{eq:data1} from the adopted model for $W_k(x)$ by plugging-in the MCML estimates. 
Fit separate standard geostatistical models as in \textit{Step 0} and compute the empirical variogram for the simulated dataset.
\item[\textit{Step 2.}] Repeat \textit{Step 1} a large enough number of times, say $M$. 
\item[\textit{Step 3.}] Use the resulting $M$ empirical variograms to generate 95\% confidence intervals at each of a set of pre-defined
distance bins.
\end{enumerate}
 If the empirical variogram in \textit{Step 0} falls fully or partly outside the $95\%$ confidence intervals, we conclude that the model is not able to capture the spatial structure of the data satisfactorily.

\section{Application I: Re-analysis of the \textit{Loa loa} data}\label{sec:application}
A total of 223 villages were sampled in the four study sites (see Web Figure 1 in Web Appendix B). The empirical prevalences from the RAPLOA and microscopy tests show a clear, approximately linear relationship when both are logit-transformed (see Web Figure 3 in Web Appendix B). Each of the two also exhibits a highly non-linear relationship with surface elevation (see Web Figure 4 in Web Appendix B), which we capture using a piecewise linear spline with knots at 750 meters and 1015 meters. 

We consider the two following models.
\begin{itemize}

\item Model 1: a slightly modified, more flexible, version of the CDRM, given by 
\begin{eqnarray}\label{eqn:CDRM2} 
\begin{cases}
{\rm logit}\{p_{1}(x_i)\} = \mu_{1}(x_i) + S_1(x_i) + Z_{i1}		\\
{\rm logit}\{p_{2}(x_i)\} = \mu_{2}(x_i)  + \alpha {\rm logit}\{p_{1}(x_i)\} + Z_{i2}
\end{cases},
\end{eqnarray}
where 
\begin{eqnarray*}
\mu_{k}(x_i) &=& \beta_{k,0}+\beta_{k,1}\min\{e(x_i), e_{1}\}+\beta_{k,2}I(e(x_i) > e_{1})\min\{ e(x_i)-e_{2}, e_{2}-e_{1}\}\\&& + \beta_{k,3}\max\{e(x_i)-e_{2},0\}, \quad k=1,2,
\end{eqnarray*}
where $e(x)$ denotes the elevation in meters at location $x$, 
$e_{1}=750$, $e_{2}=1015$ and $I(\mathcal{P})$ is
an indicator function which takes value 1 if $\mathcal{P}$ is true and 0 otherwise. In this parameterisation,
$\beta_{k,1}$ is the effect of elevation on prevalence below 750 meters, 
$\beta_{k,2}$ its effect between 760 and 1015 meters, and $\beta_{k,3}$ its effect above 1015 meters.

\item Model 2: obtained by incorporating an additional spatial process $S_{2}(x)$, independent of $S_{1}(x)$, into the linear predictor for microscopy in Model 1 to give
\begin{eqnarray}\label{eqn:CDRM3} 
{\rm logit}\{p_{2}(x_i)\} = \mu_{2}(x_i)  + S_{2}(x_{i}) + \alpha {\rm logit}\{p_{1}(x_i)\} + Z_{i2}
\end{eqnarray}
\end{itemize}

\subsection{Results}\label{sec:fit_loa}
    
Table \ref{tab:regtable_loa} reports the MCML estimates obtained for Models 1 and 2. We observe that all parameters common to both Models 1 and Model 2 have comparable point and interval estimates, except for $\tau^2_1$ which has a substantially narrower 95$\%$ confidence interval under Model 1 than Model 2. 

As expected, both models show a significant and positive logit-linear relationship between RAPLOA and miscroscopy. However, Model 2, which include the additional spatial process $S_{2}(x)$, is also able to capture spatial variation in microscopy prevalence on a scale of about 24 meters. 

We use the validation procedure of Section \ref{sec:inferences} to test which of the two models better fits the spatial structure of the data. The results (see Web Figure 5) show a satisfactory assessment of Model 2, whereas for Model 1 the empirical variogram for microscopy partly falls outside the 95$\%$ confidence band, questioning its validity.

\begin{table}[b]
 \centering
 \def\~{\hphantom{0}}
 \begin{minipage}{155mm}
  \caption{Monte Carlo maximum likelihood estimates and associated 95$\%$ confidence intervals for the fitted Model 1 and Model 2 to the \textit{Loa loa} data; see Section \ref{sec:application} for more details.} \label{tab:regtable_loa}
  \begin{tabular*}{\textwidth}
  {@{}
  l@{\extracolsep{\fill}}
  c@{\extracolsep{\fill}}
 c@{\extracolsep{\fill}}
  c@{}}
  \hline
{Parameter}         & Model 1 & Model 2 \\ \hline 
$\beta_{1, 0}$  &  -0.791&  -0.763   \\
   						&   (-1.984, 0.402) & (-1.963, 0.437) \\
$\beta_{1, 1} \times 10^3$    &0.515&   0.588  \\
   						&  (-0.977, 2.008) & (-0.922, 2.098) \\

$\beta_{1, 2} \times 10^3$  &-3.529 &  -3.412   \\
   						&(-7.314, 0.255)   & (-7.155 , 0.331) \\
$\beta_{1, 3} \times 10^3$  &-0.110 & -0.059 \\
   						& (-1.531, 1.312) &  (-1.501 , 1.382)  \\
$\beta_{2, 0}$  & -1.762 & -1.736 \\
   						&  (-2.075, -1.449)  & (-2.244, -1.229)  \\
$\beta_{2, 1} \times 10^3$    & 0.208 &  0.126  \\
   						&(-0.386, 0.802)   & (-0.799, 1.050)  \\
$\beta_{2, 2} \times 10^3$  &-0.223 &    -0.039  \\  
   						& (-2.023, 1.576)   & (-2.944 , 2.865) \\
$\beta_{2, 3} \times 10^3$  &-0.591   &  -0.612 \\
   						& (-1.666, 0.485)   & (-2.429, 1.205)  \\
$\sigma^2_{1}$ &  1.581&  1.617   \\
   						& (0.669, 3.738)    & (0.679, 3.851)  \\
$\sigma^2_{2}$ & --- &  0.216 & \\	
   						&   & (0.111, 0.419) \\			  
$\phi_{1}$ & 182.037& 187.388   \\
   						& (64.657, 512.512)    & (65.171, 538.807)  \\
$\phi_{2}$ & --- &   23.686 \\
   						&   & (6.150, 91.220)  \\
$\tau_{1}^2$ & 0.205 & 0.324\\
   						& (0.081, 0.521)   & (0.052, 6.229)  \\
$\tau_{2}^2$ & 0.324& 0.104\\ 		
   						& (0.055, 5.873)    &  (0.018, 5.797)   \\	  
$\alpha$ &1.005 & 1.017\\  
   						& (0.902, 1.107)   & (0.939, 1.095)  \\ \hline
\end{tabular*}
\end{minipage}
\vspace*{-6pt} 
\end{table}

\begin{figure}
 \centerline{\includegraphics[width=6.5in]{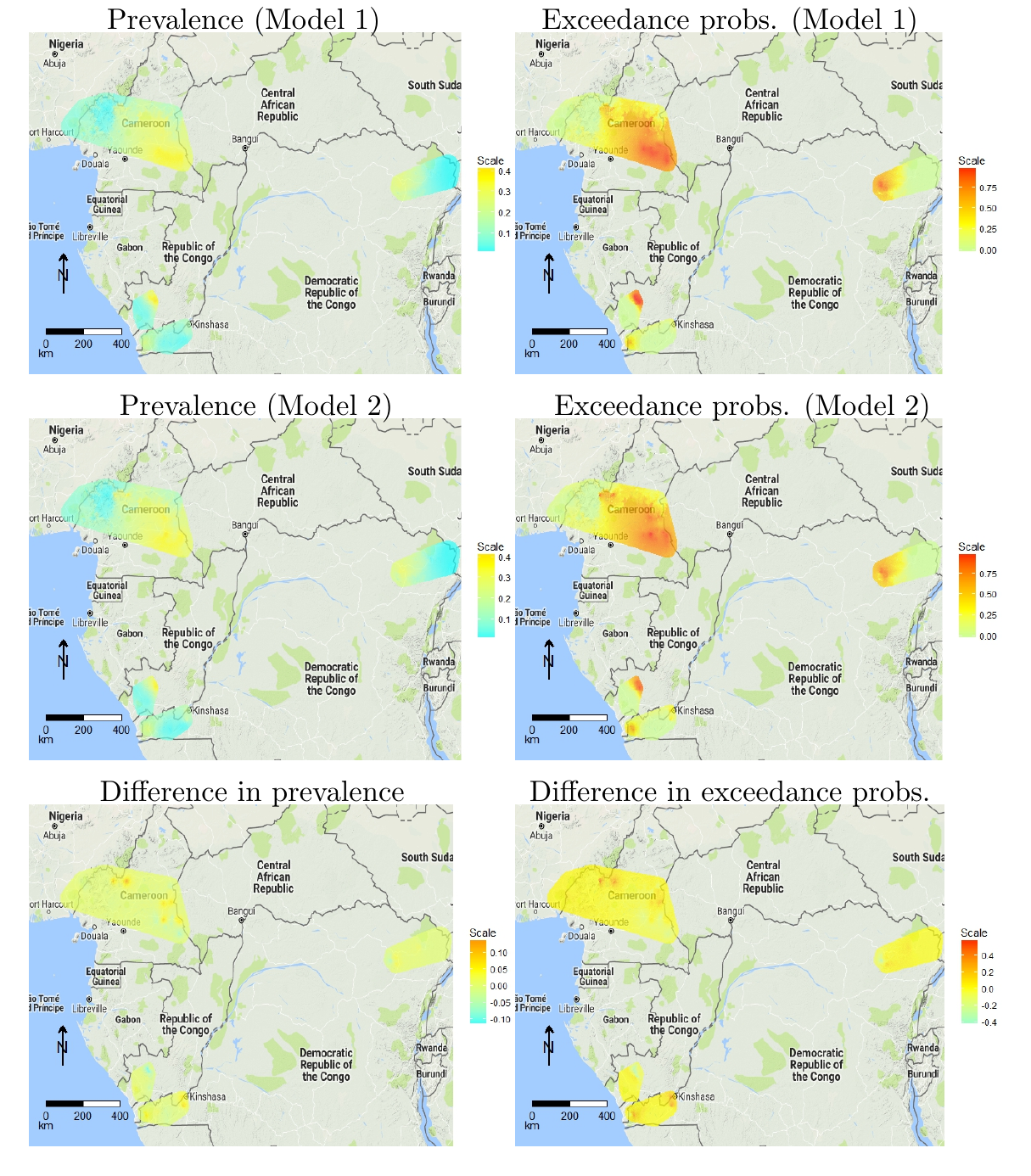}}
    \caption{Predictive mean of \textit{Loa loa} microfilariae prevalence (left panels) and probabilities of exceeding a 20$\%$ prevalence threshold (right panels), for Model 1 (top panels) and Model 2 (middle panels) of Section \ref{sec:application}. The bottom panels show the differences between the predictive surfaces of Model 2 and Model 1.}\label{fig:mean_predictions}\label{fig:relationships}
\end{figure}

We now compare the predictive inferences on microscopy prevalence between the two models in order to assess whether the introduction of $S_{2}(x)$ makes a tangible difference.
Figure \ref{fig:mean_predictions} shows the point estimates for microscopy prevalence and the exceedance probabilities for a 20$\%$ prevalence threshold under Model 1 (upper panels), under Model 2 (middle panels), and the difference between the two (lower panels). Overall, the predicted spatial pattern in prevalence from the two models is similar, but with substantial local differences. The  difference between the point estimates for prevalence ranges from -0.12 to 0.13, while the difference between the two exceedance probabilities  
ranges from -0.44 to 0.59.
 
\subsection{Simulation Study}\label{sec:simulation}
We carry out a simulation study in order to quantify the effects on the predictive inferences for prevalence when ignoring microscopy-specific residual spatial variation. To this end, we compare the predictive performances of Model 1 and Model 2 at 20 unobserved locations corresponding to the centroids of 20 clusters (shown as red points in Web Figure 1) that we identify using the k-means algorithm \citep{hartigan1979algorithm} and proceed as follows. We simulate 10,000 Binomial data-sets under Model 2 by setting its parameters to the estimates of Table \ref{tab:regtable_loa} and fit both models. We then carry out predictions for microscopy prevalence over the 20 unobserved locations. We summarise the results at each of the 20 prediction locations using  the 95$\%$ coverage probability (CP),  the root-mean-square-error (RMSE) and the 95\% predictive interval length (PIL). Table \ref{tab:simtable_loa} shows the three metrics averaged over the 20 locations for Model 1 and Model 2. The CP of Model 1 (77.5\%) is well below its nominal level of 95\%. This is also reflected by a smaller PIL for Model 1, suggesting that this provides unreliably narrow 95$\%$ predictive intervals for prevalence. Finally, we note that Model 1 also has a larger RMSE than Model 2.

\begin{table}[b]
 \vspace*{-6pt}
 \centering
 \def\~{\hphantom{0}}
  \begin{minipage}{140mm}
  \caption{Results of the simulation study including the 95$\%$ coverage probability (CP), the root-mean-square-error (RMSE), the 95\% predictive interval length (PIL) averaged over the 20 unobserved locations. For more details, see the main text in Section \ref{sec:simulation}. \label{tab:simtable_loa}}
  \begin{tabular*}
  {\columnwidth}{@{}
  l@{\extracolsep{\fill}}
  c@{\extracolsep{\fill}}
    c@{\extracolsep{\fill}}
      c@{\extracolsep{\fill}}
  c@{}}
  \hline
  & CP &   RMSE & PIL \\ \hline
Model 1 & 0.770 & 4.948  & 0.140 \\
Model 2 & 0.943 & 3.932 &  0.185\\
 \hline
\end{tabular*}\vskip18pt
\end{minipage}
\end{table}

\section{Application II: Joint prediction of \textit{Plasmodium falciparum} prevalence using RDT and PCR}\label{sec:application2}
 
The malaria data consist of 3,587 individuals sampled across 949 locations (see Web Figure 2). The outcomes from RDT ($k=1$) and PCR ($k=2$) were concordant in 92.4\% of all the individuals 
tested for \textit{P. falciparum}. This suggests that estimating 
components of
residual spatial variation that are
 unique to each diagnostic may be difficult. For this reason our model for the data takes the following form
\begin{equation}
\label{eqn:mod_mal_2}
f_k(p_{jk}(x_i)) = \beta_{k, 0} + \sum_{l=1}^3 \beta_{k, l}d_{ij,l} + \nu_k T(x_i), 
\end{equation}
where: $d_{ij,1}$ is a binary variable taking value 1 if the $j$-th individual at $x_{i}$ is a male and 0 otherwise; $d_{ij,2}=\min\{a_{ij},5\}$ and $d_{ij, 3}= \max\{a_{ij}-5, 0\}$, i.e. the effect of age, $a_{ij}$,
is modelled as a linear spline with a knot at 5 years. 

\subsection{Results}

Table \ref{tab:regtable_mal} reports point estimates and 95\% confidence intervals for the model parameters. Gender has a significant effect on PCR prevalence, but its effect on RDT prevalence is not significant at the conventional 5\% alpha level. The effect of age is comparable between the two diagnostics, with the probability of a positive test increasing with age up to 5 years and decreasing thereafter. The estimated variance component, $\hat{\nu}_{1}^2=0.230$, associated with RDT is about three times that for PCR, $\hat{\nu}_{2}^2=0.081$. The spatial process $T(x)$, common to both diagnostics, accounts for spatial variation in malaria prevalence up to a scale of about 11.6 kilometers, beyond which the correlation falls below 0.05. The variogram-based validation procedure of Section \label{sec:inferences} does not show any strong evidence against the fitted model (see Web Figure 6 in Web Appendix B).

\begin{table}[b]
 \centering
 \def\~{\hphantom{0}}
 \begin{minipage}{140mm}
  \caption{Monte Carlo maximum likelihood estimates and associated 95$\%$ confidence intervals for the model in \eqref{eqn:mod_mal_2} fitted to the malaria data. \label{tab:regtable_mal}}
  \begin{tabular*}{\textwidth}
  {@{}
  l@{\extracolsep{\fill}}
  c@{\extracolsep{\fill}}
  c@{\extracolsep{\fill}}
  c@{}}
  \hline
{Parameter}         & RDT ($k=1$) & PCR ($k=2$) \\ \hline
$\beta_{k, 0}$&  -6.186   & -4.373   \\ 
& (-7.234, -5.138) & (-17.008, 8.261) \\
$\beta_{k, 1}$  &    -0.003  &    0.251   \\ 
& (-0.415, 0.395) & (0.009, 0.494) \\
$\beta_{k, 2}$  &   0.261 &  0.220 \\ 
& (0.070, 0.453) &  (0.095, 0.344) \\
$\beta_{k, 3}$   &  -0.059  & -0.020  \\ 
& (-0.081, -0.037) & (-0.028, -0.012) \\
$\nu_k^2$ &   0.230    & 0.081   \\ 
 & (0.145, 0.364) &  (0.052, 0.126) \\
 \hline
$\phi_T$  & 11.581 \\
&  (10.618, 12.63)   \\
\hline
\end{tabular*}
\end{minipage}
\vspace*{-6pt}
\end{table}

\begin{figure}
   \centering
   \centerline{\includegraphics[width=4.4in]{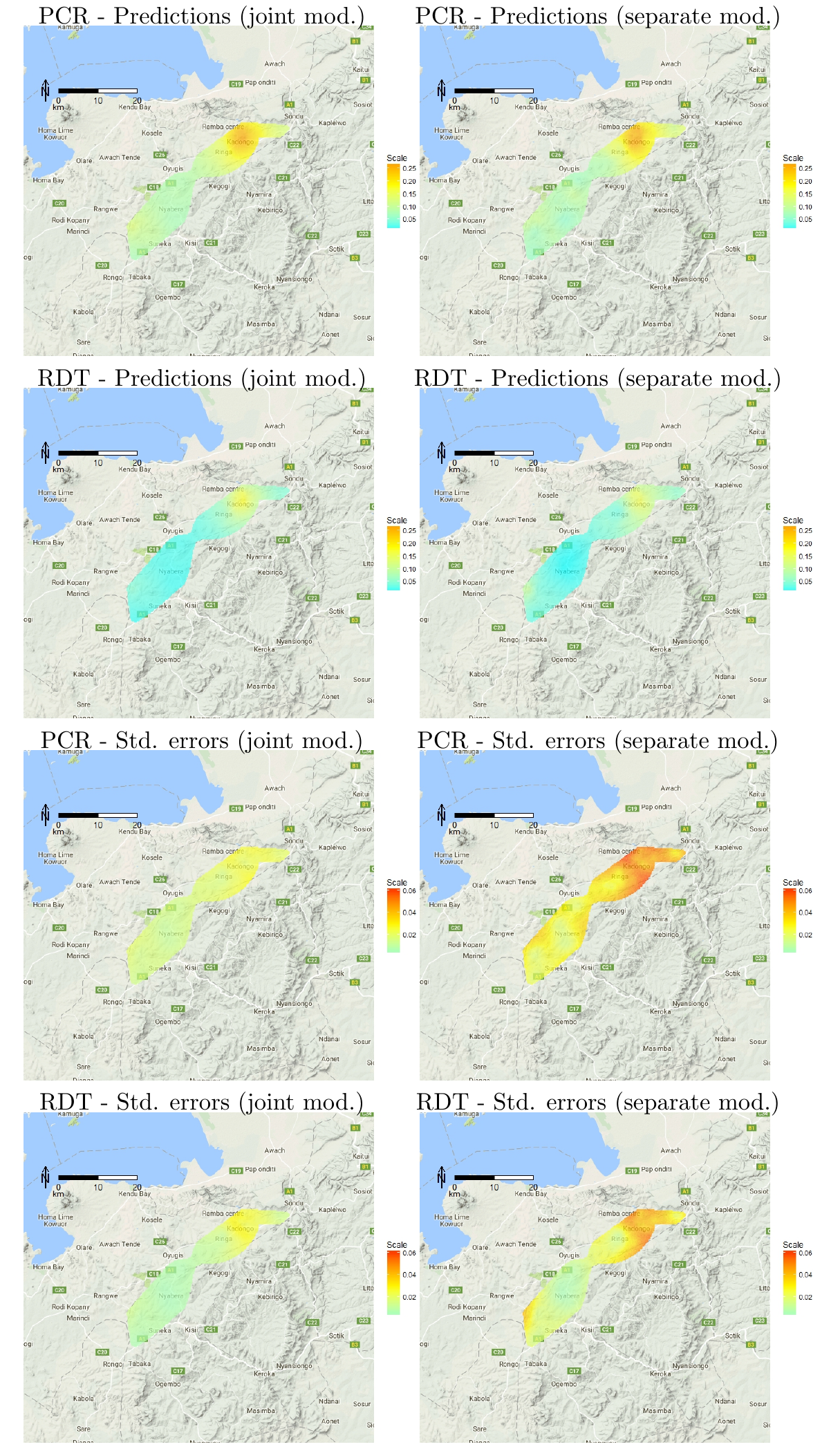}}
    \caption{Point predictions (first and second rows) and standard errors (third and fourth rows) of \textit{P. falciparum} prevalence for five-year-old children under the joint geostatistical model in \eqref{eqn:mod_mal_2} (left panels) and two separate geostatistical models (right panels) for RDT and PCR prevalence. }\label{fig:mean_pred_mal}
\end{figure}

To quantify the benefit of carrying out a joint analysis for RDT and PCR, we compare the predictive inferences for prevalence that are obtained under two scenarios: ($i$) the fitted model in  \eqref{eqn:mod_mal_2};  ($ii$)  separate fitted models that ignore the cross-correlation between the outcomes of the two diagnostic tests. Figure \ref{fig:mean_pred_mal} shows the point predictions and standard errors for RDT and PCR prevalences for five-year-old male children under scenarios ($i$) (left panels) and ($ii$) (right panels). We observe that the point predictions for prevalence under the two models are strongly similar but the joint model in \eqref{eqn:mod_mal_2}, as expected, yields smaller standard errors throughout the study area.

Having chosen \eqref{eqn:mod_mal_2} as the best model, we compare the exceedance probabilities (EPs) for a 10$\%$ threshold between RDT and PCR. Using each of the two diagnostics, we then identify malaria 
hotspots, as the sets of locations such that their EP is 
at least 90$\%$. Figure 7 of Web Appendix B shows that PCR identifies a considerably larger hotspot in the north east of the study area than
does RDT, and a smaller hotspot in the south west 
that is undetected by RDT. These results are consistent with the main findings of \citet{mogeni2017detecting}.

\section{Conclusions and extensions}\label{sec:discussion}

We have developed a flexible geostatistical framework to model reported disease counts from multiple diagnostics and have distinguished two main classes of problems: (1) prediction of prevalence as defined
by a gold-standard diagnostic using data obtained from a
more feasible low-cost, but potentially
biased, alternative; (2) joint prediction of prevalences as defined
by two diagnostic tests. As the burden of disease declines in endemic regions, the use of multiple transmission metrics and diagnostics becomes necessary in order to better inform and adapt control strategies. It is thus important to develop suitable methods of inference that allow the borrowing of strength of information across multiple diagnostics. As our study has shown, the main benefit of this approach is 
a reduction in the uncertainty associated with the predictive inferences on disease risk.
 
Our application to Loiasis mapping has shown the importance of acknowledging the existence of
residual spatial variation specific to each diagnostic test. Through a simulation study, we have also shown that ignoring this source of extra-Binomial variation can lead to unreliably narrow prediction intervals for prevalence, with actual coverages falling well below their nominal level. 

The second application on malaria mapping has highlighted the benefits of a joint analysis of data from two diagnostic tests when both are of scientific interest. A joint model can yield estimates of prevalence with smaller standard errors than estimates
obtained from two separate geostatistical models. 

Although we have only considered the case of two diagnostic tests throughout the paper, our methodology can be easily extended to more than two. However, the nature of the extension
 will be dependent on the specific context and scientific goal. For example, a natural extension of the models of Section \ref{sec:model_formulation1} would be to use multiple biased diagnostic tools (for $k=1,\ldots,K-1$) to better predict a gold-standard ($k=K$). In this case, the cross-correlation between the outcomes of the biased diagnostic tests could be modelled using the symmetric structure of the model in Section \ref{sec:model_formulation2}, while preserving an asymmetric form for the linear predictor of the gold-standard. Formally, this is expressed as
\begin{eqnarray}\label{eqn:mod_ext}
\begin{cases}
f_k\{p_{jk}(x_i)\} = d_{ijk}^\top \beta_k + \nu_k\left[ S_k(x_i) + T(x_i) \right] + Z_{ik}, \quad  k=1, \dots , K-1\\
f_{K}\{p_{jK}(x_i)\} = d_{ijK}^\top\beta_K + S_{K}(x_i) + Z_{iK} +\sum\limits_{k=1}^{K-1} \alpha_k f_k\{p_j(x_i)\}
\end{cases}.
\end{eqnarray}
However, we would be wary of attempting to fit this, or other comparably
complex models without an initial exploratory analysis that might help to understand the extent of the cross-correlations between the outcomes of different diagnostics,
with a view to reducing the dimensionality of the model.


\section*{Acknowledgements}
BA holds an Economic and Social Research Council North-West Doctoral Training Centre funded doctoral studentship (1619934).

The {\it Loa loa} data were collected by  field and lab teams led by S. Wanji, M.N. Mutro, and F. Tepage with financial support from the UNICEF/UNDP/World Bank/WHO Special Programme for Research and Training in Tropical Diseases and the African Programme for Onchocerciasis Control \citep{wanji2012validation}.

The authors would like to thank Dr Gillian Stresman from the London School of Hygiene and Tropical Medicine and Dr Jennifer Stevenson from the Johns Hopkins Bloomberg School of Public Health for providing the data on \textit{Plasmodium falciparum} and for useful discussions on the epidemiological aspects of the study.



\section*{Supplementary Materials}

Web Appendix A, referenced in Section \ref{sec:inferences}; and Web Appendix B, referenced in Sections \ref{sec:problem}, \ref{sec:application}, and \ref{sec:application2} are available with this paper at the Biometrics website on Wiley Online Library.


%

\bibliographystyle{biom} \bibliography{refs}





\label{lastpage}

\end{document}